# Electronic structure of monolayers of group V atoms: Puckering and spin orbit interaction in nano-slabs

*Dat T. Do\* and Subhendra D. Mahanti\**

Department of Physics and Astronomy, Michigan State University, East Lansing, MI 48824

Abstract

Inspired by the observation of mass less Dirac particles associated with low energy excitations of 2-dimensional (2D) honeycomb (HC) lattices of group IV atoms (C, Si, Ge, Sn) near the Dirac points (DPs) and spin-orbit interaction (SOI) induced gaps at the DPs, we have investigated the electronic structure of 2D HC lattices of group V atoms (As, Sb, Bi). Here also one sees DPs at the K points in the Brillouin zone for both planar (flat) and puckered sheets. Unlike the group IV systems the Fermi energy in group V systems lies above the DPs. The flat sheets are metallic but undergo structural distortions to form puckered sheets that are semiconducting. SOI profoundly alters the band structure, opens up gaps at the DPs, and in binary systems BiSb and SbAs gives large Rashba-type spin splitting. Nano-slabs of group V atoms show excellent thermoelectric properties, particularly in the hole-doped regime.

Graphene has been one of the most exciting areas of physics during the last decade. [1,2] Its low energy physics in the absence of spin orbit interaction (SOI) is governed by the massless Dirac particles associated with the linearly dispersing electronic band structure centered around the K and K' points of the 2-dimensional (2D) Brillouin zone. Similar features are present in other group IV ($s^2p^2$) systems, such as silecene, germanene, and tinene. [3] SOI opens a gap at the K (K') point making the massless Dirac quasi-particles massive. In contrast, physical properties of 2D group V ($s^2p^3$) atoms have not been explored in great detail. There are several physical motivations to study theoretically the electronic structure and related properties of monolayers of these atoms (Bi, Sb, As) and their binary mixtures. Since in 3D, As, Sb, and Bi are semimetals, [4] have distinctly different atomic structures from their group IV counterparts partly driven by Peierl's distortion physics, and exhibit different $sp^3$ bonding characteristics (lone pairs in group V systems), one would like to know how these properties change in 2D and how they differ from graphene, silecene etc. The other important questions are (i) whether in 2D these systems are semimetals or semiconductors and (ii) what is the role of structural distortions (puckering) of 2D layers on the electronic properties. One would also like to know the electronic transport properties of these group V nanoslabs.

To address these questions we have investigated the electronic structures of single layers of As, Sb, Bi, SbAs and BiSb, focusing on four aspects: (i) semimetal vs semiconductor, (ii) Dirac point physics, (iii) effect of puckering on band structure and, (iv) role of SOI. Details of calculations are given in the supplemental document. [5] The atomic structure of a monolayer looking straight down is shown in **Figure 1** (top panel) in which A and B denote different sites in a HC lattice. Looking sideways, we have considered two structures, one flat (like graphene) and the other puckered (**Figure 1** (bottom panel)). When we consider the mixed systems, i.e. SbAs or BiSb, A and B atoms correspond to the two types of atoms.

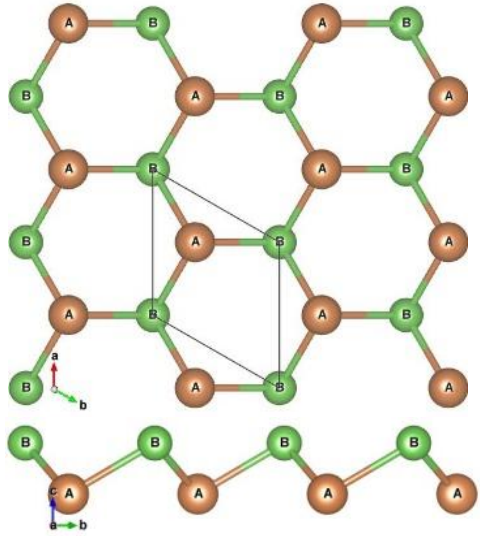

**Figure 1:** Top view (upper) and side view (lower) of a single puckered layer of group V atoms. A and B indicate different sites in the honeycomb lattice.

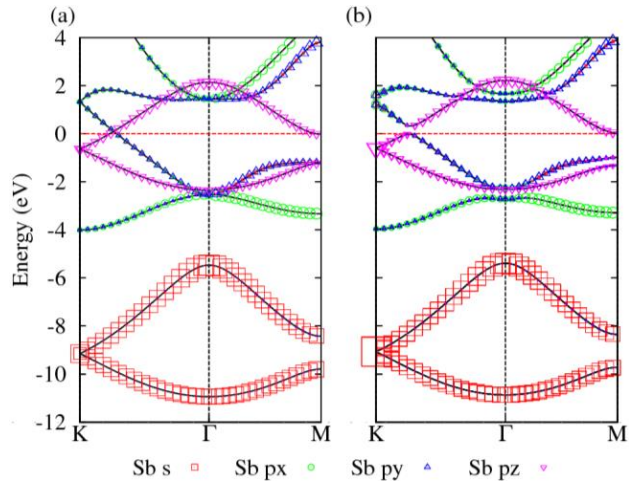

**Figure 2:** Band structure of a flat Sb nano-slab without (a) and with (b) spin-orbit interaction

Since the general features of the band structures are common to As, Sb, and Bi we will discuss the results for one (Sb) in detail. The band structures along K→Γ→M without and with SOI for a flat slab are shown in **Figure 2** a,b. The two *s*-bands lie between -6 and -11 eV and show DPs at the K points, deep below the Fermi energy (FE). The $p_z$ bands do not mix with the $p_x$ and $p_y$ bands due to symmetry and have band structures similar to the s-bands. In the absence of SOI, they also show a DP near the K point. This $p_z$-band DP is responsible for all the interesting physics in graphene because the Fermi energy (FE) goes right through this DP. However in Sb due to the extra electron/atom the FE is about 0.5 eV above this DP. As in graphene the $p_x$ and $p_y$ bands are degenerate at the $\Gamma$ point but mix as one moves away from this point. One of the $p_x$-$p_y$ bonding bands and the bonding $p_z$ band lie below the FE and are completely occupied. The Fermi energy cuts the second $p_x$-$p_y$ bonding bands and the antibonding $p_z$ band along the $\Gamma K$ direction. These bands however lie below the Fermi energy along the $\Gamma M$ direction. Along $\Gamma K$ there is band crossing between these two bands right at the FE; we denote this crossing point as $\mathbf{k_0}$. When SOI is turned on, a small gap (~50 meV) appears at the upper DP and a large gap (~200 meV) appears at $\mathbf{k_0}$. SOC also removes the $p_x$-$p_y$ degeneracy at the $\Gamma$ point.

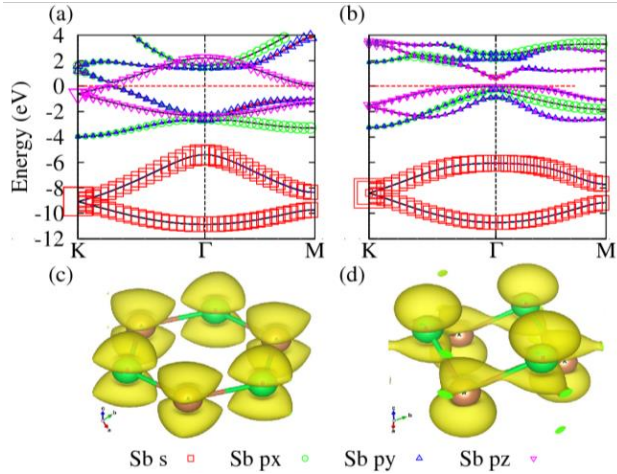

**Figure 3:** Band structures including SOI of planar (a) and puckered (b) layers of Sb. Charge densities associated with bonding $p_z$ at $\Gamma$ for the flat (c) and the puckered sheet (d).

When the Sb atoms are allowed to pucker similar to what is seen in 3D Sb, 4 i.e. when nearest neighbor Sb atoms are allowed to displace upward and downward perpendicular to the flat layer respectively, the energy decreases by ~0.47 eV/atom. Puckering is accompanied by a drastic rearrangement of the p-bands and the system becomes a semiconductor with a direct band gap of 0.69 eV at the Γ point. In **Figure 3** we give the band structures of the planar and puckered layers of Sb and the charge density associated with the bonding $p_z$ orbital at the Γ point (details of the metal-insulator transition as one goes from planar to puckered structure is given in supplemental document [5]). Valence band is very flat near Γ (large hole effective mass) suggesting that the nano-slab has the potential for a good 2D thermoelectric. The SOI induced gap at the upper DP changes from 49 meV for the flat geometry and is 172 meV for the puckered geometry, thus SOI induced mass gets enhanced by puckering.

To see the effect of puckering on the wave functions we have plotted the charge densities associated with the bonding $p_z$ state at the Γ point for the flat layer and the valence band maximum for the puckered layer, also dominated by the $p_z$ state. For the latter, there is a thin sheet of bonding charge density in the central plane which controls the charge transport through holes near the valence band maximum. The thermopower associated with these highly confined holes is discussed below.

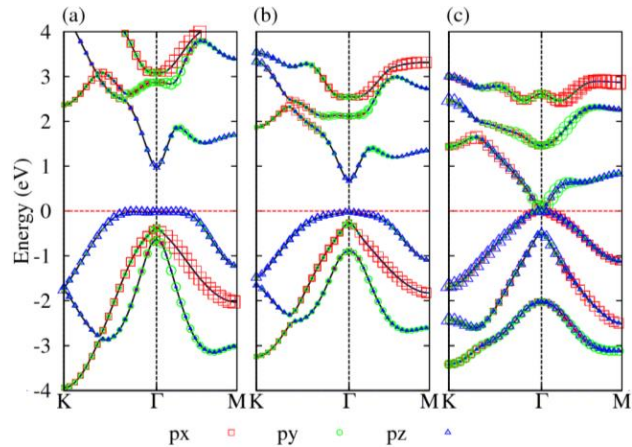

**Figure 4:** Band structures including SOI of puckered layers of (a) As (Eg=0.98 eV), (b) Sb (Eg=0.69 eV), and (c) Bi (Eg=0.095 eV).

**Figure 4** shows how the band structure changes from As (weak SOI) to Sb (intermediate SOI) to Bi (strong SOC) for the puckered geometry. The band gaps decrease from As→Sb→Bi, both with and without SOI. We find SOI reduces the band gap as seen in bulk systems PbTe, $Bi_2Te_3$ etc, the reduction is very small for As, from 0.99 eV to 0.98 eV, intermediate for Sb, from 0.85 to 0.69 and quite large for Bi, from 0.46 eV to 0.095 eV. The band gap for Bi is small ~0.1 eV. Also for Bi, the effect of SOI is so strong that $p_z$ state gets mixed with $p_x$ and $p_y$, and as a result, the top of valence band become a mixture of $p_z$ and $p_x$ while the lowest of the conduction band has strong $p_y$ character. However GGA (which is used here) gaps tend to be small and an improved method will give larger band gaps. The other important observation is the stronger effect of SOI on the DP associated with pz orbitals due to puckering. The SOI induced gap is ~0.001 eV for As, ~0.1 eV for Sb and ~1 eV for Bi. Thus we conclude that both puckering and SOI play significant roles in the electronic structure of group V atom nano-slabs, similar to what is seen in C, Si, Ge nano-slabs.[3]

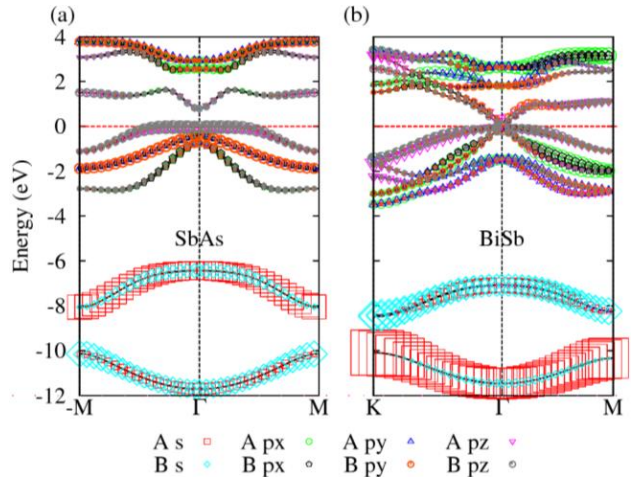

**Figure 5:** Band structures including SOI for puckered layers of (a) SbAs ($E_g$=0.81 eV) and (b) SbBi ($E_g$=0.19 eV). A and B indicate Sb and As for SbAs and Bi and Sb for BiSb respectively.

In addition to the pure systems we studied ordered binary alloys (SbAs and BiSb) to see how mixing affects the band structure, particularly the effect of SOI. **Figure 5** a,b gives the band for SbAs and BiSb respectively. The band gaps lie between the respective pure systems. The compounds show Rashba-type spin splitting [6], the splitting being very large for SbBi.

The other interesting feature is the splitting of the s-bands at the DP. In a simple nearest neighbor single-band tight binding, the two sublattices of a HC lattice decouple at the K point (origin of the DP). Thus the two energy levels should correspond to the corresponding atomic energies. For SbAs the order is normal, As s-state is below the Sb s-state by ~2 eV, the atomic energies being -18.92 (-16.03) eV for As (Sb). In contrast, the ordering of Sb and Bi is anomalous, Bi s-state is lower than the Sb s-state by ~2 eV, the atomic energy for Bi is -15.19 eV, higher than that for Sb. One can understand this anomaly by incorporating intra-sublattice hopping.

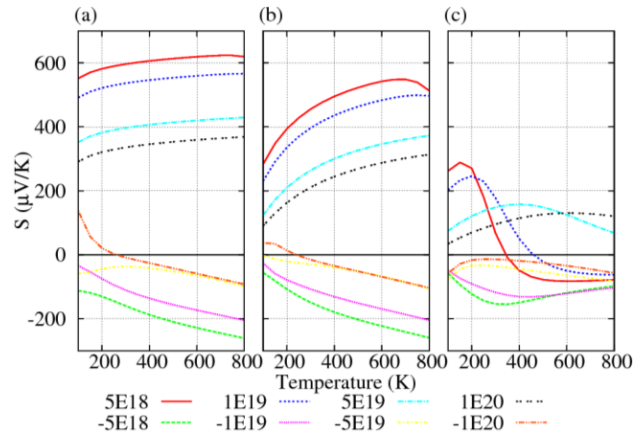

**Figure 6:** Calculated thermopower of (a) As, (b) Sb, and (c) Bi, spin-orbit coupling is included.

Since the nano-slabs are narrow band-gap semiconductors they are potentially good thermoelctrics (TE). [7] Looking at the top most valence band and the lowest conduction band we expect them to be excellent *p*-type thermoelectric due to large hole effective mass. To test this we have calculated the thermopower *S* using Boltzmann transport equation in constant relaxation time approximation.[8] In **Figure 6** we give $S$ for As, Sb, Bi both for electron and hole dopings. The peak values of |$S$| are very large, ~500 – 600 μV/K for holes (*p*-type) for Sb and As and ~300 μV/K for electrons (*n*-type). The Bi is not a great *p*-type TE due mainly to large bipolar effect (which is the cause of $S$ changing from positive to negative values) at higher temperatures caused by small band gaps.

ASSOCIATED CONTENT
**Supporting Information**. Computational details and additional figures. This material is available free of charge via the Internet at http://pubs.acs.org.

AUTHOR INFORMATION

**Corresponding Authors**

*E-mail: dodat@msu.edu; mahanti@pa.msu.edu

ACKNOWLEDGMENT

This work was supported by the Center for Revolutionary Materials for Solid State Energy Conversion, an Energy Frontier Research Center funded by the U.S. Department of Energy, Office of Science, Office of Basic Energy Sciences under Award Number DE-SC0001054.

REFERENCES

1. Geim, A. K.; Novoselov, K. S. The rise of graphene. *Nat. Mater.* **2007,** *6* (3), 183--191.

2. Allen, M. J.; Tung, V. C.; Kaner, R. B. Honeycomb Carbon: A Review of Graphene. *Chem. Rev.* **2010,** *110* (1), 132--145.

3. Matthes, L.; Pulci, O.; Bechstedt, F. Massive Dirac quasiparticles in the optical absorbance of graphene, silicene, germanene, and tinene. *J. Phys.: Condens. Matter* **2013,** *25* (39), 395305.

4. Shoemaker, D. P.; Chasapis, T. C.; Do, D.; Francisco, M. C.; Chung, D. y.; Mahanti, S. D.; Llobet, A.; Kanatzidis, M. G. Chemical ordering rather than random alloying in SbAs. *Phys. Rev. B* **2013,** *87* (9), 094201.

5. Supplemental Documents. ACS. http://pubs.acs.org.

6. Rashba, E. I. *Sov. Phys. Solid Sate* **1960,** *2,* 1109.

7. Sofo, J. O.; Mahan, G. D. Optimum band gap of a thermoelectric material. *Phys. Rev. B* **1994,** *49* (7), 4564--4570.

8. Madsen, G. K. H.; Singh, D. J. BoltzTraP. A code for calculating band-structure dependent quantities. *Comput. Phys. Commun.* **2006,** *175* (1), 67--71.

# Supplemental Document:

# "Electronic structure of monolayers of group V atoms: Puckering and spin orbit interaction in nano-slabs"

*Dat T. Do and Subhendra D. Mahanti*

Department of Physics and Astronomy, Michigan State University, East Lansing, MI 48824

**Computational details**

We model the nano-slabs using a supercell method where slabs of A where (A=Bi,Sb,As) or BiSb, SbAs are separated by a thick vacuum layer (>10Å). All systems are subjected to constrained relaxation according to the physical situations under study. We employ density functional theory using pseudo-potential and projector augmented wave (PAW) [1,2] method with generalized gradient approximation (GGA) as implemented in Vienna Atomic Simulation Package (VASP). [3,4,5] For the exchange-correlation potential, we used the parameterized model developed by Perdew, Burke and Ernzerhoff (PBE) [6]. The total energy calculations were done with 16 × 16 × 1 Monkhorst-Pack k-mesh [7], energy cutoff of 400 eV and convergence criteria of 1E-4 eV.

The thermopower calculations were carried out using Boltzmann's transport theory in the relaxation time ($\tau$) approximation. [8] For simplicity we neglect the energy dependence of $\tau$. In order to calculate thermopower, we first calculate band structures for a dense k-mesh of 46 × 46 × 1 using Wien2k [9] and then apply BoltzTraP code [8].

**Additional figures**

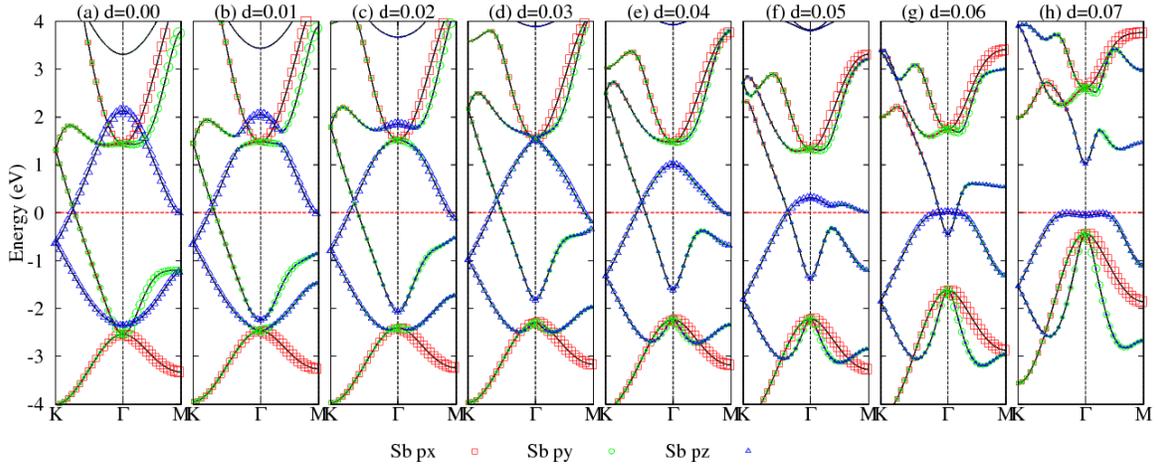

**Figure 1:** Changes in the band structure of a single layer of Sb as it is subjected to puckering distortion where the two inequivalent atoms ($Sb_1$=A and $Sb_2$=B, see Fig. 1 of the text) of a hexagonal lattice are allowed to displace in opposite directions perpendicular to the hexagon plane. *d* (in Å) is a measure of the relative displacement of A and B in *z* direction. The metal to semimetal transition takes place around d=0.06 Å$^o$ and semimetal to semiconductor transition takes place around d=0.7Å$^o$ which is equal to the value in bulk Sb.

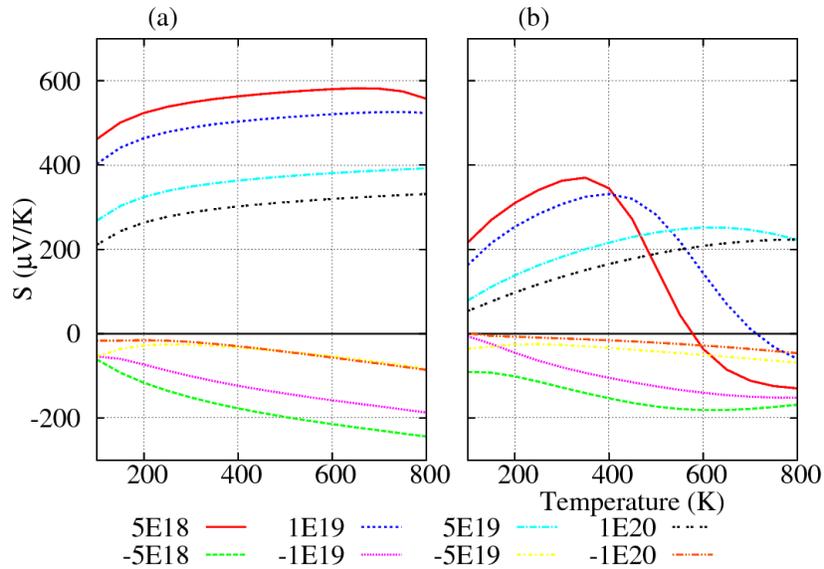

**Figure 2:** Thermopower of (a) SbAs and (b) BiSb calculated using Boltzmann's transport formalism in constant relaxation time approximation. The p-doped SbAs shows large

thermopower over a broad temperature range. The rapid drop in p-type thrmopower in BiSb is due to smaller band gap which enhances electronic contribution at higher temperature due to thermal excitation (bipolar effect).

REFERENCES

1. Blochl, P. E. Projector augmented-wave method. *Phys. Rev. B* **1994,** *50,* 17953--17979.

2. Kresse, G.; Joubert, D. From ultrasoft pseudopotentials to the projector augmented-wave method. *Phys. Rev. B* **1999,** *59* (3), 1758001775.

3. Kresse, G.; Hafner, J. Ab initio molecular dynamics for liquid metals. *Phys. Rev. B* **1993,** *47* (1), 558--561.

4. Kresse, G.; Furthmuller, J. Efficiency of ab-initio total energy calculations for metals and semiconductors using a plane-wave basis set. *Comput. Mater. Sci.* **1996,** *6* (1), 15--50.

5. Kresse, G.; Furthmuller, J. Efficient iterative schemes for ab initio total energy calculations using a plane-wave basis set. *Phys. Rev. B* **1996,** *54* (16), 11169--11186.

6. Perdew, J. P.; Burke, K.; Ernzerhof, M. Generalized gradient approximation made simple. *Phys. Rev. Lett.* **1996,** *77* (18), 3865--3868.

7. Monkhorst, H. J.; Pack, J. D. Special points for Brillouin-zone integrations. *Phys. Rev. B* **1976,** *13,* 13.

8. Madsen, G. K. H.; Singh, D. J. BoltzTraP. A code for calculating band-structure dependent quantities. *Comput. Phys. Commun.* **2006,** *175* (1), 67--71.

9. Blaha, P.; Schwarz, K.; Madsen, G. K. H.; Kvasnicka, D.; Luitz, J. *WIEN2k, An Augmented Plane Wave + Local Orbitals Program for Calculating Crystal Properties;* Karlheinz Schwarz, Techn. Universitat Wien: Austria, 2001.